\documentclass[prb,showpacs]{revtex4}
\usepackage{graphicx,psfrag,amssymb,amsmath}
\begin{document}
\bibliographystyle{apsrev}
\title{Imperfect nesting and
transport properties in unconventional density waves}
\author{Bal\'azs D\'ora}
\affiliation{Department of Physics, Budapest University of Technology and 
Economics, H-1521 Budapest, Hungary}
\author{Kazumi Maki}
\affiliation{Department of Physics and Astronomy, University of Southern
California, Los Angeles CA 90089-0484, USA}
\affiliation{Max Planck Institute for the Physics of Complex Systems, N\"othnitzer Str. 38, D-01187, Dresden, Germany}
\author{Attila Virosztek}
\affiliation{Department of Physics, Budapest University of Technology and 
Economics, H-1521 Budapest, Hungary}
\affiliation{Research Institute for Solid State Physics and Optics, P.O.Box
49, H-1525 Budapest, Hungary}

\date{\today}

\begin{abstract}
We consider the effect of imperfect nesting in quasi-one dimensional unconventional density
waves.
 The phase diagram is very close to those
in a conventional DW. 
The linear and non-linear aspects of the electric conductivity are
discussed. At $T=0$ the frequency dependent electric conductivity develops 
a small dip at low frequencies.
 The threshold electric field depends strongly on the imperfect
nesting parameter, allowing us to describe the measured threshold electric
field in the low temperature phase of the quasi-two dimensional organic conductor,
$\alpha$-(BEDT-TTF)$_2$KHg(SCN)$_4$ very well.   
\end{abstract}

\pacs{75.30.Fv, 71.45.Lr, 72.15.Eb, 72.15.Nj}

\maketitle

\section{Introduction}

Unconventional density waves (i.e. density waves with momentum dependent
gap) have long been suspected to be the possible
ground state of certain materials. Since  the original proposal\cite{HR}, 
a number of studies have been done to discover the properties of these
systems under various conditions: in two dimensions, the effect of magnetic
field was investigated\cite{Ners1,Ners2}, and the stability of the
different phases was studied\cite{Schulz}. In three dimensions, the phase
diagram and the thermodynamic properties were discussed \cite{GG}.
But certainly one of the main issues is the ``rediscovery'' of
unconventional charge density waves with
higher angular momentum, the so-called d-density wave ($\Delta({\bf k})\sim
\cos(k_xa)-\cos(k_ya)$), which was investigated
exhaustively \cite{nayak,carbotte} over the past few years, and seems to
convincingly describe 
many aspects
of underdoped high $T_c$ superconductors\cite{benfatto}. In this exotic
phase,
the ground state is characterized by circulating electric currents around
the plaquettes of the lattice, inducing some kind of magnetic order. This
is why this phase was previously known as orbital
antiferromagnet\cite{Ozaki}.

Recently we have reported about different aspects of unconventional spin
density waves (USDW) and unconventional charge density waves
(UCDW)\cite{nagycikk,rapid,tesla,cap,impur} in quasi-one dimensional
systems. 
In analogy with unconventional 
superconductivity\cite{szupravezetes,revmod}, the extension of the
conventional DW theory looks very natural to the case where the gap
depends on the quasiparticle momentum like $\Delta({\bf k})\sim\sin(ck_z)$. In Ref.
\onlinecite{nagycikk}, a detailed
description of the thermodynamics and optical conductivity of USDW is
given. Interestingly, the UDW phase transition is not accompanied by a
spatially periodic modulation of the spin or charge density. In Ref. 
\onlinecite{rapid}, we studied the threshold electric field in perfectly nested
UDW. In Ref. \onlinecite{tesla,cap}, the effect of applied magnetic field is discussed, focusing on
the phase diagram and the threshold electric field. In
Ref. \onlinecite{impur}, the effect of impurities were treated in the Born limit. 

On the other hand, the nature of the low temperature phase (LTP) of the 
$\alpha$-(ET)$_2$KHg(SCN)$_4$ is not yet
fully understood. In the  LTP,
there is no X-ray or NMR signal\cite{nmr1,nmr2} characteristic to conventional DW \cite{jetp,singl}, which
naturally follows from our theory\cite{nagycikk}. The destruction of the LTP in increasing
magnetic field suggests that it is a kind of a CDW\cite{montambo1,montambo2,tesla}, and the experimentally
obtained phase diagram\cite{jetp} is very close to our theoretical prediction in the
Pauli limiting case in UCDW\cite{tesla,cap}.

Recently, the threshold electric field associated with the sliding motion
of DW of $\alpha$-(ET)$_2$KHg(SCN)$_4$ has been reported\cite{ltp,sasaki}. It shows strong
temperature dependence contrary to the one in conventional SDW, while it is
not divergent at $T_c$ as one would expect from a conventional CDW\cite{viro1,viro2}. Our
previous analysis on the threshold electric field shows good agreement
with the experimental data especially at low temperatures\cite{rapid,tesla}. The strong, monotonic
enhancement of the threshold
electric field in an applied
magnetic field\cite{sasaki} also coincides with our theory.   

The Fermi surface of the $\alpha$-(ET)$_2$ salts contains large quasi-one
dimensional sheets favouring the density wave phase transition, and small
two dimensional pockets\cite{fermi}. Since the quasi-one dimensional sheets
do not necessarily nest each other perfectly, it looks necessary to consider
theoretically imperfectly nested systems to improve the matching between theory and
experiment. As in conventional DW\cite{huang1,huang2}, this effect can be handled 
in terms of imperfect nesting. In a recent paper, we succeeded in describing
the measured angle dependent magnetoresistance data in $\alpha$-(ET)$_2$
salt in the presence of imperfect nesting\cite{admr}. This is why we
believe that the inclusion of imperfect nesting can help us to describe the
threshold electric field as well.

The object of the present paper is to extend the earlier analysis\cite{rapid,tesla} in the
presence of imperfect nesting. We discuss the temperature
dependence of the order parameter at different $\epsilon_0$'s, where $\epsilon_0$
is the imperfect nesting parameter defined by the strongly anisotropic nearest
neighbor tight-binding spectrum $\varepsilon(k)$
and by the nesting vector $\bf Q$ as 
$\varepsilon({\bf k})+\varepsilon({\bf k-Q})=2\epsilon_0\cos(2bk_y)$. The phase boundary is
almost the same as in a conventional
DW. 
In the density of states the peaks
at $\pm\Delta$ split into cusps at $\pm\Delta\pm\epsilon_0$. 
In the optical conductivity, only the low frequency part is modified due to imperfect
nesting. The weight of the 
Dirac delta peak at zero frequency is finite at all temperatures. 
The threshold electric
field depends strongly on $\epsilon_0$, and at $\epsilon_0=0.8\Delta_{00}$
$E_T$ shows very good agreement with the experimental data on
$\alpha$-(ET)$_2$KHg(SCN)$_4$\cite{ltp}. Therefore we may conclude that the LTP in
$\alpha$-(ET)$_2$ salts is most likely an imperfectly nested UCDW. Some 
preliminary results have already been published in Ref. \onlinecite{physicaB}.

\section{Phase diagram}

\begin{figure}[h!]
\psfrag{y}[b][t][1][0]{$T/T_{c0}$, $\Delta(0,\epsilon_0)/\Delta_{00}$}
\psfrag{x}[t][b][1][0]{$\epsilon_0/\Delta_{00}$}
{\includegraphics[width=13cm,height=8cm]{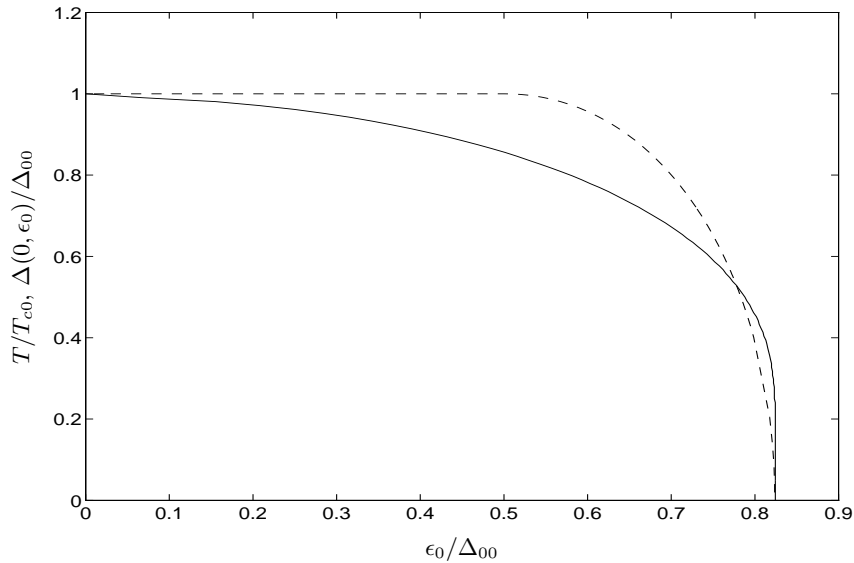}}
\caption{The phase diagram (solid line) and the zero temperature order
parameter (dashed line) are
 plotted in the presence of imperfect nesting.\label{fig:fazis} }
\end{figure}

\begin{figure}[h!]
\psfrag{x}[t][b][1][0]{$T/T_{c0}$}
\psfrag{y}[b][t][1][0]{$\Delta(T,\epsilon_0)/\Delta_{00}$}
{\includegraphics[width=13cm,height=8cm]{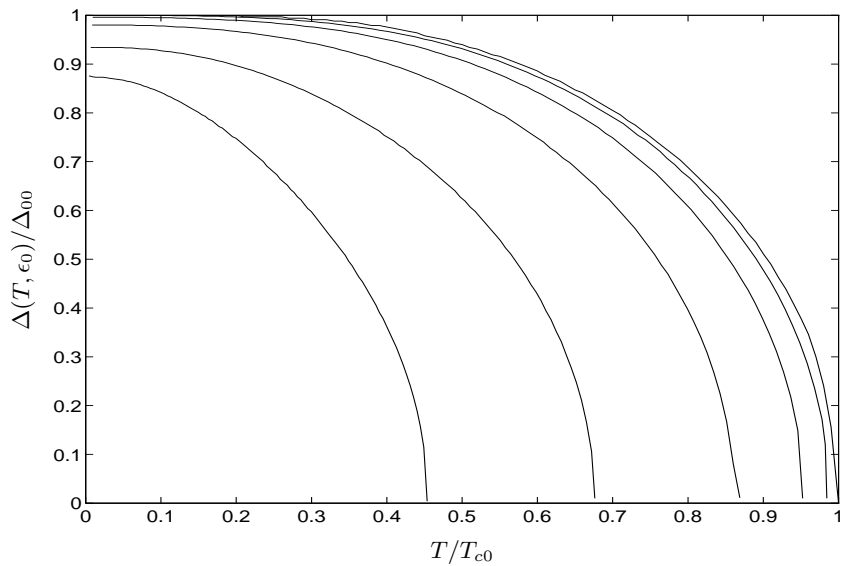}}
\caption{The temperature dependence of
the order parameter is shown for $\epsilon_0/\Delta_{00}=0$,
$0.1$, $0.3$, $0.5$, $0.7$ and $0.8$ from right to left.\label{fig:delta3d}}
\end{figure}

As a model we take the highly anisotropic quasi-one dimensional tight-binding
Hamiltonian for the kinetic energy, and consider the interaction between the on-site
and nearest neighbor electrons on orthogonal lattice\cite{nagycikk}. The
gap depends on the wavevector like $\Delta\sin(ck_z)$ or $\Delta\cos(ck_z)$.
The effect of imperfect nesting is incorporated in the theory by replacing
the Matsubara frequency with
$\omega_n+i\epsilon_0\cos(2bk_y)$\cite{huang1,huang2}.
With this, the gap-equation reads as
\begin{equation}
1=T\pi P
\frac{N_0}{4}\sum_n\int_0^{2\pi}\int_0^{2\pi}\frac{\sin(z)^2dzdy}{\sqrt{(\omega_n+i\epsilon_0\cos(2y))^2
+\Delta^2\sin(z)^2}},
\end{equation}
where $P$ is the interaction responsible for the UDW formation, $N_0$ is
the density of states in the normal state at the Fermi energy per spin.
From this, by taking $\Delta=0$, the second order phase boundary is given by
\begin{equation}
1=T\pi P \frac{N_0}{4}\sum_n\frac{1}{\sqrt{\omega_n^2+\epsilon_0^2}}.
\end{equation}
This equation is identical to the BCS gap-equation for s-wave
superconductor or conventional DW after replacing
$\epsilon_0$ with $\Delta_0$, the order parameter of a perfect nested
conventional DW\cite{nagycikk}. 
From this, the second order phase boundary is given by $\epsilon_0=\Delta_0(T_c)$,
$\Delta_0(T)$ is the temperature dependence of the gap in a perfect nested
conventional DW with $T_{c0}$ transition temperature.
 The critical nesting
is given by $\epsilon_0=\sqrt e\Delta_{00}/2\approx0.82\Delta_{00}$, where
$\Delta_{00}$ is the gap in a perfectly nested system at zero
temperature. At $T=0$ the order parameter remains unchanged for
$2\epsilon_0<\Delta_{00}$, then
vanishes rapidly.
This together with the phase diagram is shown in Fig. \ref{fig:fazis}. 
The gap-equation was solved numerically for arbitrary temperatures, and the
 results are shown in Fig. \ref{fig:delta3d}.
 
\section{Density of states}
 
The density of states is given by
\begin{equation}
N(E)=N_0\frac{4}{\pi^2}\int\limits_0^{\pi/2}\int\limits_0^{\pi/2}
\textmd{Re}\frac{|E-\epsilon_0\cos(2y)|}{\sqrt{(E-\epsilon_0\cos(2y))^2-\Delta^2
\sin^2(z)}}dydz=\frac{1}{\pi}\int\limits_0^\pi \rho(E-\epsilon_0\cos y)dy,
\end{equation}
where $\rho(E)$ is the density of states in the perfectly nested 
system\cite{nagycikk} and is the same as those in a d-wave
superconductor\cite{szupravezetes}.

\begin{figure}[h!]
\psfrag{x}[t][b][1][0]{$E/\Delta$}
\psfrag{y}[b][t][1][0]{$N(E)/N_0$}
{\includegraphics[width=13cm,height=8cm]{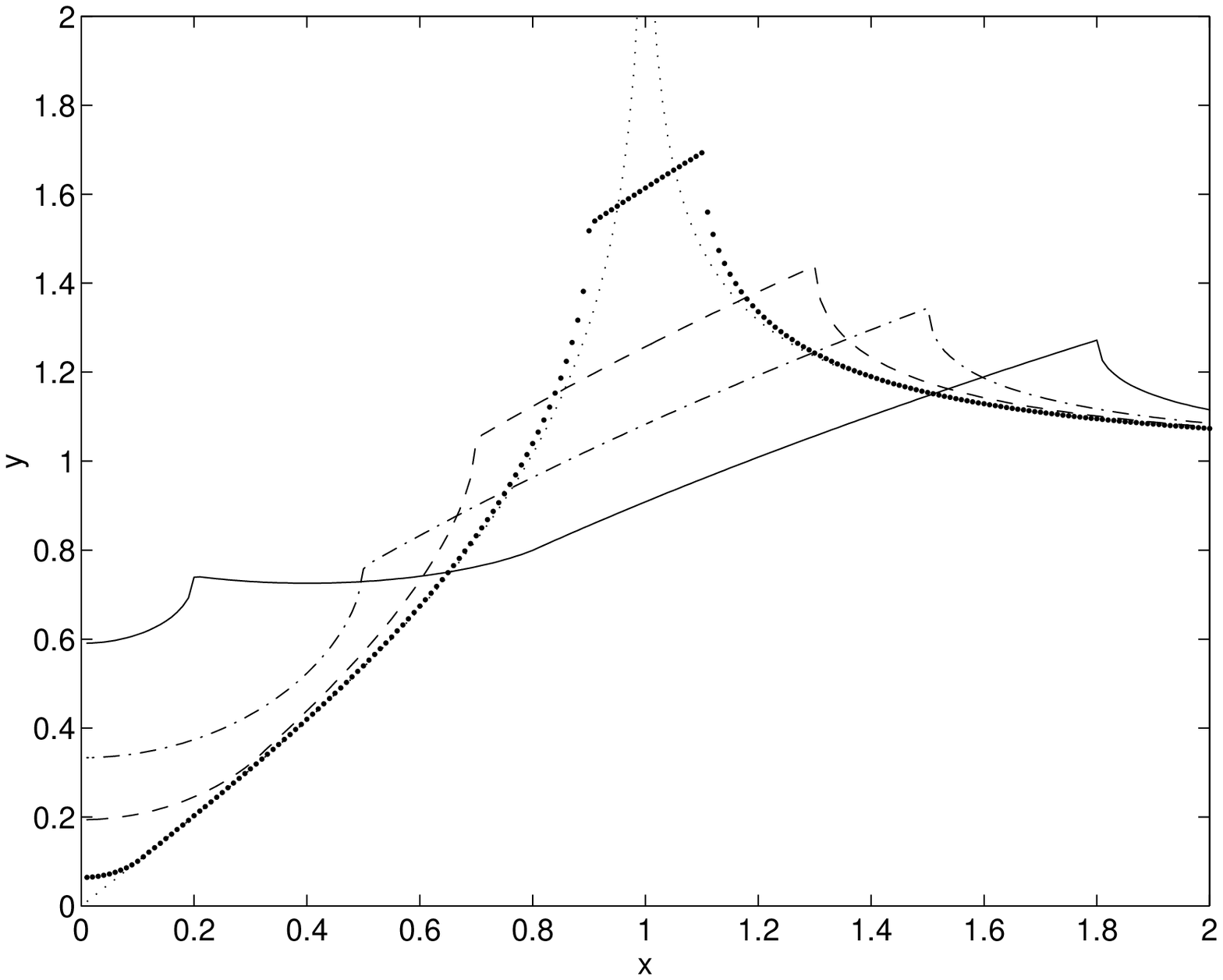}}
\psfrag{0.8}[t][l][0.4][0]{$0.8$}
\psfrag{1}[t][l][0.4][0]{$1$}
\psfrag{0.6}[t][l][0.4][0]{$0.6$}
\psfrag{0.4}[t][l][0.4][0]{$0.4$}
\psfrag{0.2}[t][l][0.4][0]{$0.2$}
\psfrag{0}[t][l][0.4][0]{$0$}
\psfrag{x}[t][b][0.7][0]{$\epsilon_0/\Delta$}
\psfrag{y}[b][t][0.7][0]{$N(0)/N_0$}
\hspace*{-5.15cm}{\includegraphics[width=3.4cm,height=4cm]{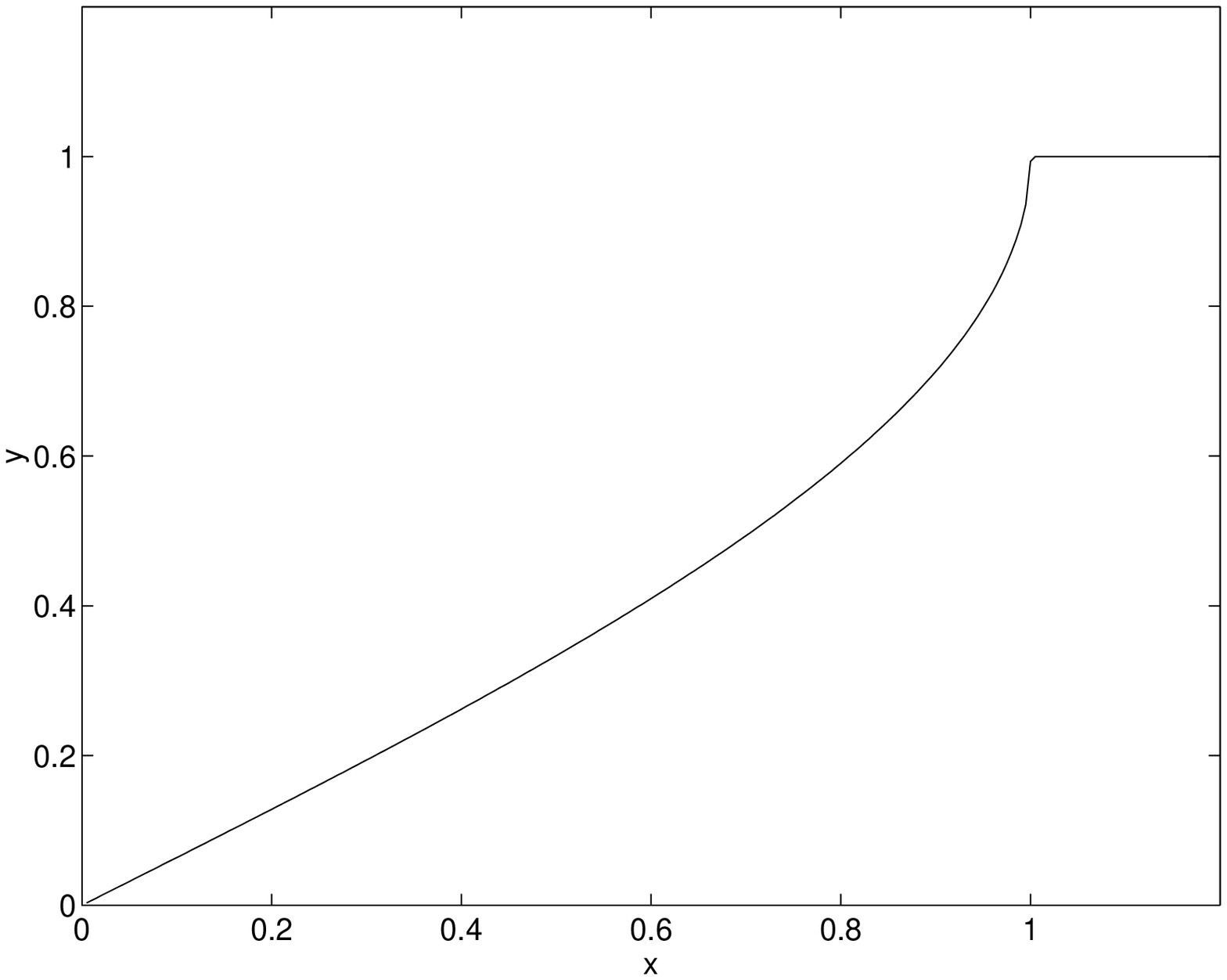}}
\caption{The quasiparticle density of states is plotted as a function of 
energy for $\epsilon_0/\Delta=0$ (thin dotted line), $0.1$ (thick dotted line),
 $0.3$ (dashed line), $0.5$ (dashed-dotted line) and $0.8$ (solid line). The inset
shows the residual density of states as a function of $\epsilon_0/\Delta$.\label{fig:dos3d}} 
\end{figure}  

In Fig. \ref{fig:dos3d}, we show $N(E)/N_0$ for $\epsilon_0/\Delta=0$, $0.1$,
$0.3$, $0.5$ and $0.8$. The inset shows the residual density of states, which
is obtained as 
\begin{equation}
N(0)=N_0\frac 2\pi \arcsin\frac{\epsilon_0}{\Delta}.\label{eq:res3d}
\end{equation}
 The peaks at $\pm\Delta$ split into cusps at $\pm\Delta\pm\epsilon_0$. 
As it is readily seen from Fig. \ref{fig:dos3d} the DOS provides
a clear signature of UDW with imperfect nesting. 
As a consequence of the finite number of occupied states at the Fermi energy,
the specific heat increases linearly with temperature close to $T=0$:
\begin{equation}
C_V(T)=N(0)\frac{2\pi^2}{3}k_B^2 T,
\end{equation}
which is in accordance with the measured specific heat in the
$\alpha-$(ET)$_2$ salts\cite{fajho}.

\section{Optical conductivity}

\begin{figure}[h!]
\psfrag{x}[t][b][1][0]{$\omega/\Delta_{00}$}
\psfrag{y}[b][t][1][0]{Re$\sigma_{yy}(\omega)\Delta_{00}/e^2N_0 v_y^2$}
{\includegraphics[width=13cm,height=8cm]{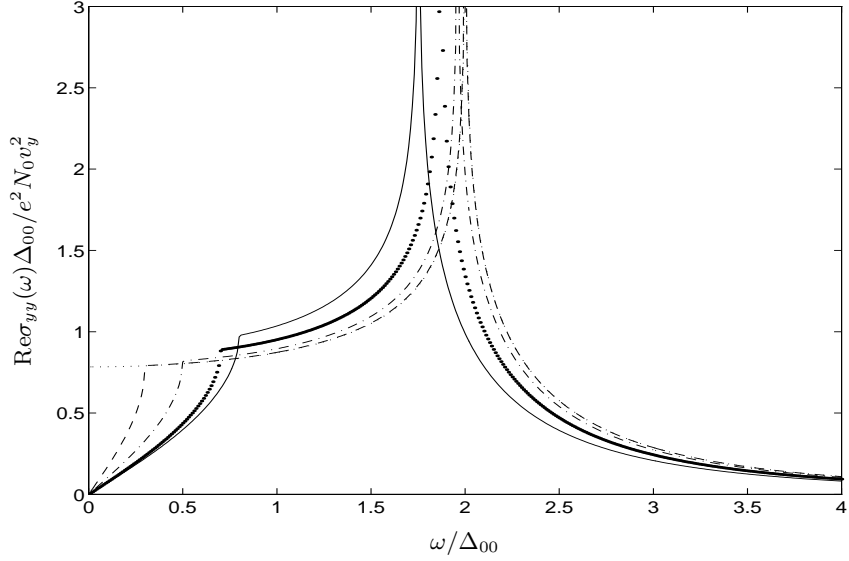}}
\caption{The optical conductivity 
in the $y$ direction is shown 
for $\epsilon_0/\Delta(0,\epsilon_0)=0$ (thin dotted line), $0.3$ (dashed line), 
$0.5$ (dashed-dotted line), $0.7$ (thick dotted line) and $0.8$ (solid line).\label{fig:vx3d}}
\end{figure}

\begin{figure}[h!]
\psfrag{x}[t][b][1][0]{$\omega/\Delta_{00}$}
\psfrag{y}[b][t][1][0]{Re$\sigma_{zz}^{sin}(\omega)2\Delta_{00}/e^2N_0
v_z^2$}
{\includegraphics[width=13cm,height=8cm]{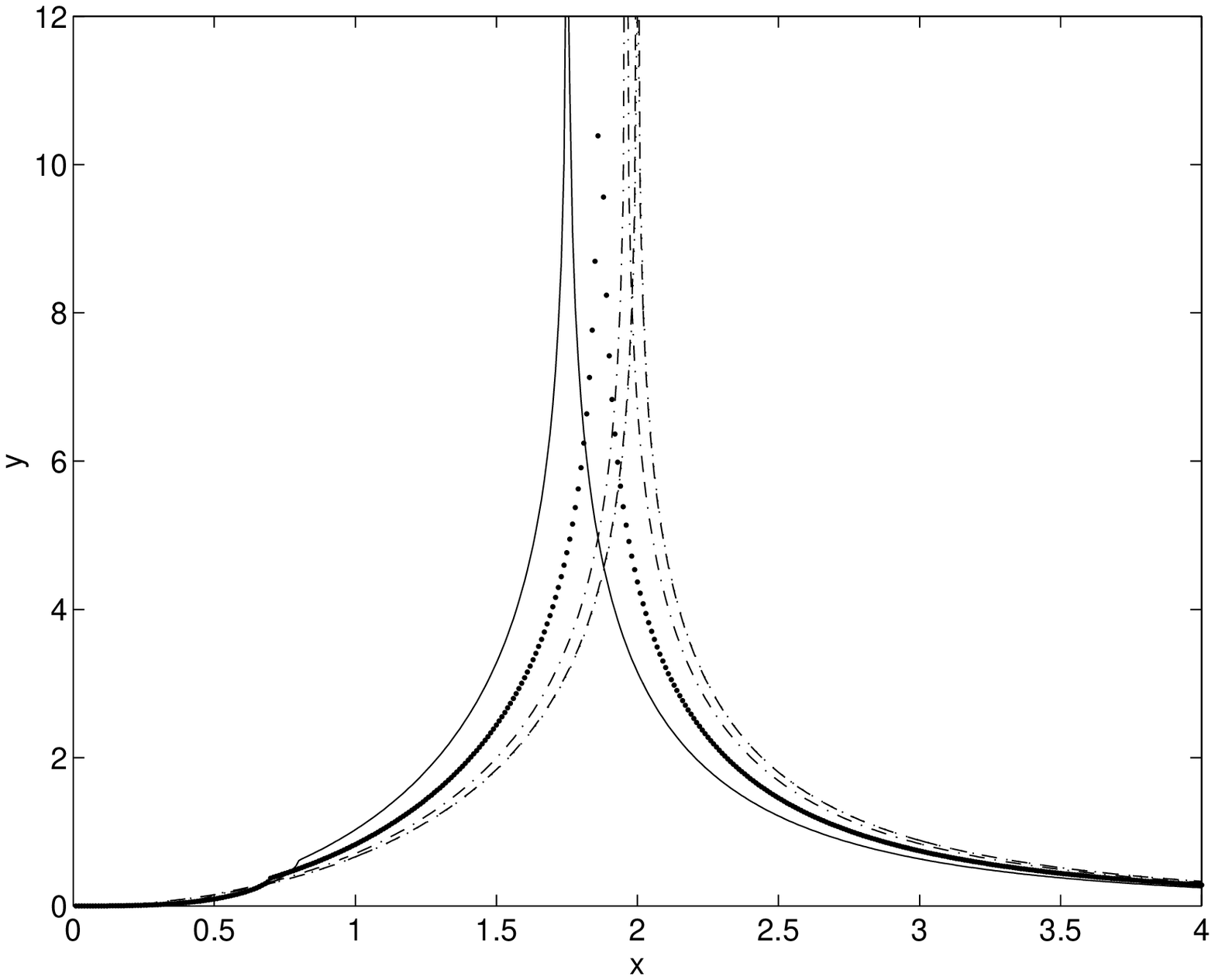}}
\caption{The optical conductivity for a
sinusoidal gap in the $z$ direction is shown 
for $\epsilon_0/\Delta(0,\epsilon_0)=0$ (thin dotted line), $0.3$ (dashed line), 
$0.5$ (dashed-dotted line), $0.7$ (thick dotted line) and $0.8$ (solid line).\label{fig:vys3d}}
\end{figure}

\begin{figure}[h!]
\psfrag{x}[t][b][1][0]{$\omega/\Delta_{00}$}
\psfrag{y}[b][t][1][0]{Re$\sigma_{zz}^{cos}(\omega)2\Delta_{00}/e^2N_0
v_z^2$}
{\includegraphics[width=13cm,height=8cm]{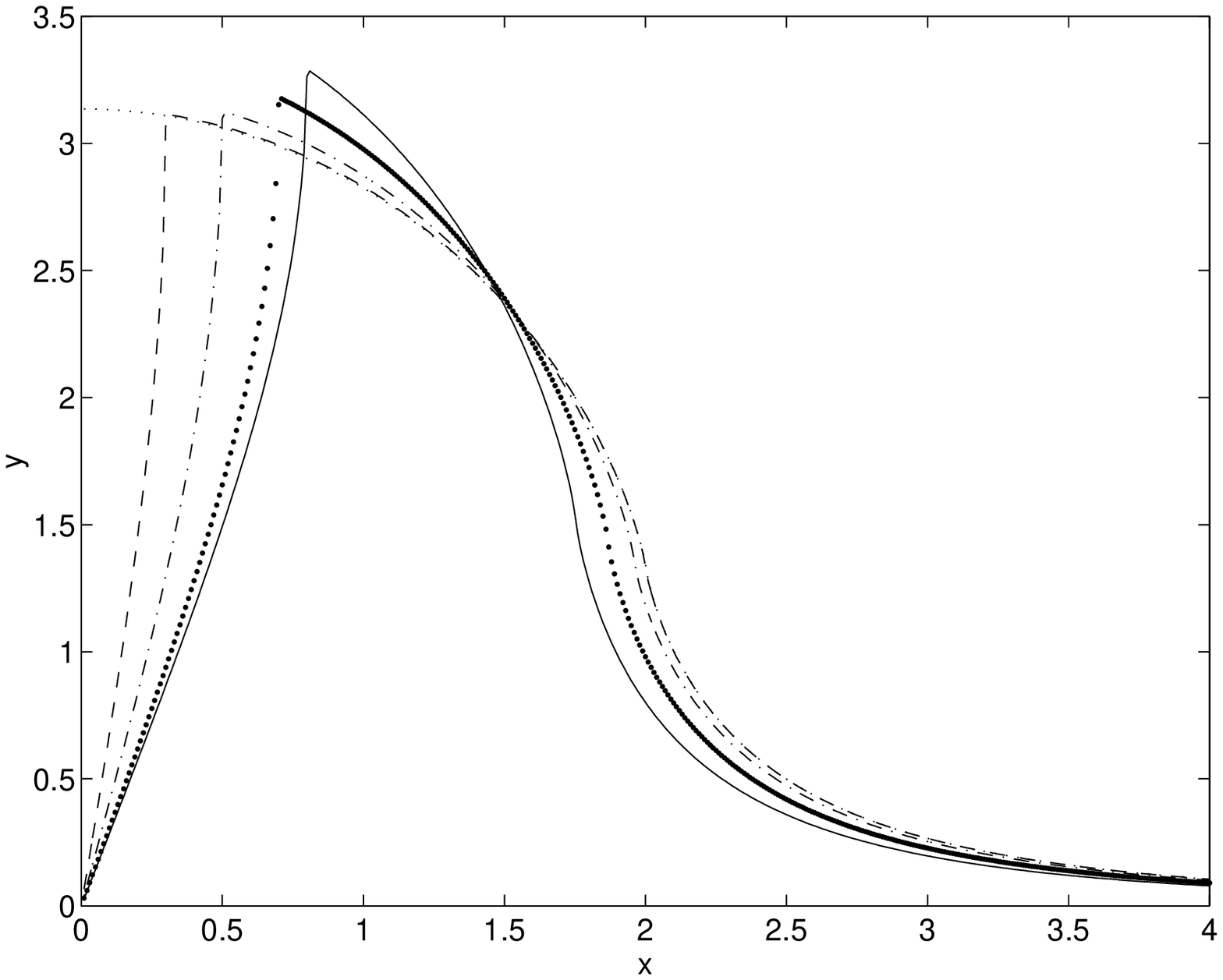}}
\caption{The optical conductivity for a
cosinusoidal gap in the $z$ direction is shown 
for $\epsilon_0/\Delta(0,\epsilon_0)=0$ (thin dotted line), $0.3$ (dashed line), 
$0.5$ (dashed-dotted line), $0.7$ (thick dotted line) and $0.8$ (solid line).\label{fig:vyc3d}}
\end{figure}

In this section we investigate the quasiparticle contribution to the
optical conductivity. For simplicity we neglect the effect of the quasiparticle 
damping due to impurity scattering
for example. The quasiparticle part
of the conductivity contains relevant information about the system
in the perpendicular cases ($y$ and $z$) when the effect of the collective contributions
can be neglected. The regular part of the optical conductivity is given by
\begin{equation}
\textmd{Re}\sigma_{\alpha\beta}^{reg}(\omega)=N_0\frac{\pi e^2}{2\omega^2}\int_{-\pi}^{\pi}\frac{d(bk_y)}{2\pi}\int_{-\pi}^{\pi}\frac{d(ck_z)}{2\pi}\textmd{Re}\frac{v_\alpha({\bf
k})v_\beta({\bf k})\Delta^2({\bf k})}{\sqrt{(\omega/2)^2-\Delta^2({\bf
k})}}\left(\tanh\left(\frac{|\omega|-2\eta}{4T}\right)+\tanh\left(\frac{|\omega|+2\eta}{4T}\right)\right),
\end{equation}
where $v_\alpha({\bf k})$ is the quasiparticle velocity in the $\alpha$
direction, $\eta=\epsilon_0\cos(2bk_y)$. From now on we restrict our
investigation to the $T=0$K case.
The optical conductivity remains
unchanged for $\omega>2\epsilon_0$ and is given by 
\begin{equation}
\textmd{Re}\sigma(\omega,\epsilon_0)=\textmd{Re}\sigma(\omega,0)\frac 2\pi
\arcsin\left(\frac{\omega}{2\epsilon_0}\right)
\end{equation}
in the remaining interval. 
The optical conductivity is shown for electric field applied in the $y$ 
direction in
Fig. \ref{fig:vx3d}. The same curve belongs to both a sinusoidal or cosinusoidal 
gap function. In Figs. \ref{fig:vys3d} and \ref{fig:vyc3d},
the optical conductivity is plotted in the $z$ direction for gap function 
$\Delta({\bf k})=\Delta\sin(ck_z)$ and $\Delta({\bf k})=\Delta\cos(ck_z)$, 
respectively. 
In this case, when the maximums of the velocity ($\sim\sin(ck_z)$) and
of the gap coincide,  divergent peak shows up at $2\Delta$. 
On the other hand, when the position of the gap maximum matches the zeros of 
the velocity,
these excitations cannot produce any divergence.
At first sight the sum rule seems to be violated since a lot of optical
weight is missing at small
frequencies. But the $\delta(\omega)$ part of
the conductivity does not freeze out at $T\rightarrow 0$ in the presence of imperfect nesting,
and its coefficient provides the missing area. 

So far the optical conductivity of $\alpha$-(ET)$_2$KHg(SCN)$_4$ salt has
not been investigated comprehensively, only very few data are available at
the moment\cite{dressel}. For more decisive statements, more experiments
are required especially in the low temperature regime.

\section{Threshold electric field}

\begin{figure}[h!]
\psfrag{x}[t][b][1][0]{$T/T_{c0}$}
\psfrag{y}[b][t][1][0]{$\rho_s(T,\epsilon_0)/\rho_s(0,0)$}
\vspace*{0cm}
{\includegraphics[width=13cm,height=8cm]{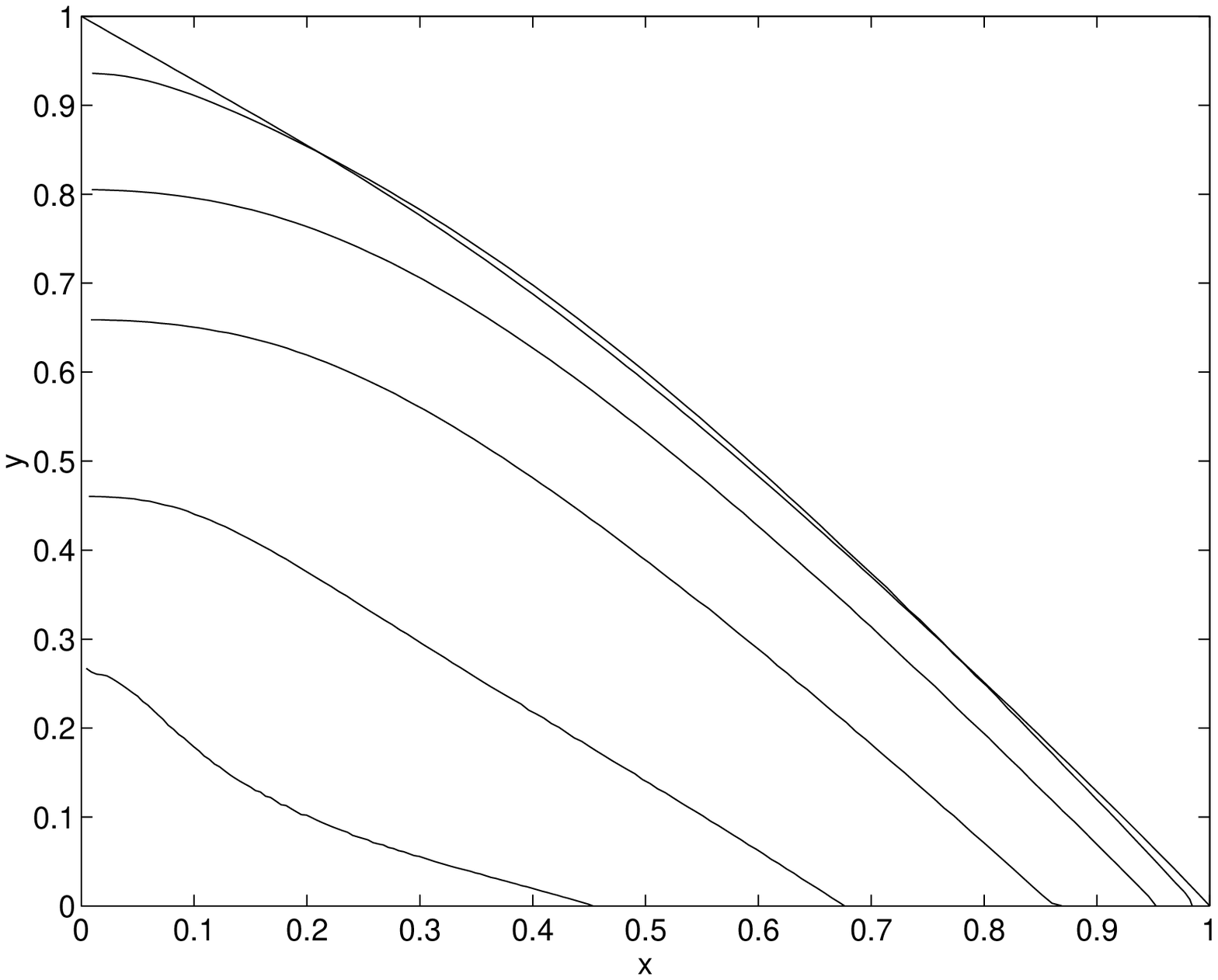}}
\caption{The condensate density is shown as a function of temperature
for $\epsilon_0/\Delta_{00}=0$,
$0.1$, $0.3$, $0.5$, $0.7$ and $0.8$ from top to bottom. \label{fig:sfd3d}}
\end{figure}

The behaviour of the non-linear conductivity is also of great interest.  
In terms of the phase $\Phi({\bf r},t)$ of DW the phase Hamiltonian is given by \cite{viro1,viro2}
\begin{eqnarray}
H(\Phi)={\bf \int}d^3r\left\{\frac 1 4 N_0 f \left[ v_F^2 \left(\frac{\partial\Phi} {\partial x}\right)^2
+v_b^2 \left(\frac{\partial\Phi} {\partial y}\right)^2
+v_c^2 \left(\frac{\partial\Phi} {\partial
z}\right)^2+
\left(\frac{\partial\Phi}
{\partial t}\right)^2-4v_FeE\Phi\right]+V_{imp}(\Phi)\right\} \label{phaseH}
\end{eqnarray}
where $f=\rho_s(T,\epsilon_0)/\rho_s(0,0)$, $\rho_s(T,\epsilon_0)$ is the condensate 
density and $E$ is an electric field applied in the $x$ direction. 
Here $v_F$, $v_b$ and $v_c$ are the characteristic velocities of the
quasi-one dimensional electron system in the three spatial directions. 
The condensate density is evaluated from the well-known formula:
\begin{equation}
\rho_s=1-\frac{1}{4T}\int_{-\infty}^{\infty}dE\frac{N(E)}{N_0}
\textmd{sech}^2\frac{E}{2T}.
\end{equation}
 At arbitrary temperatures it is evaluated numerically and shown in Fig. 
\ref{fig:sfd3d} for a set of $\epsilon_0$.

\begin{figure}[h!]
\psfrag{x}[t][b][1][0]{$T/T_{c0}$}
\psfrag{y}[b][t][1][0]{$E_T^S(T,\epsilon_0)/E_T^S(0,0)$}
{\includegraphics[width=13cm,height=8cm]{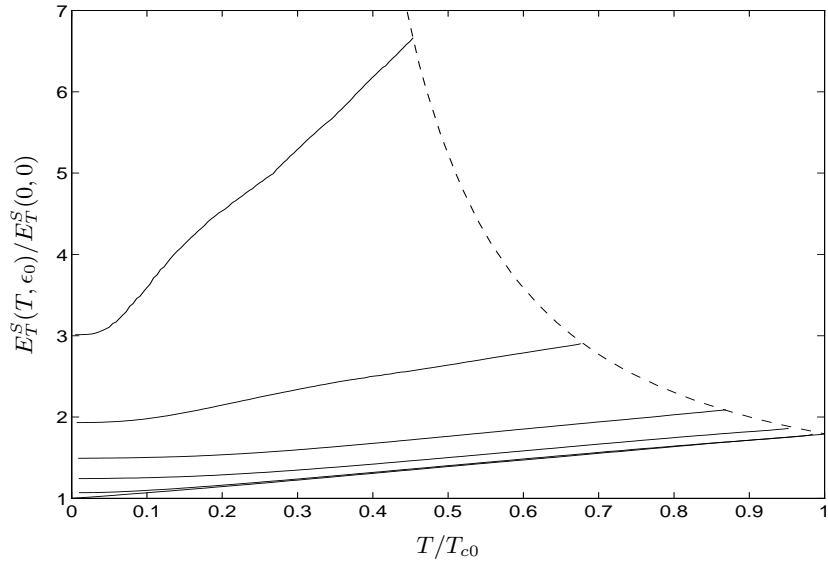}}
\caption{The threshold electric field in the strong pinning limit is plotted for 
as a function of temperature for
$\epsilon_0/\Delta_{00}=0$, $0.1$, $0.3$, $0.5$, $0.7$ and $0.8$ from bottom to
top. The dashed line represents $E_T$ along the phase boundary.\label{fig:et3d}}
\end{figure}

Assuming a nonlocal impurity potential as in Refs. \onlinecite{rapid,tesla}  
\begin{equation}
U({\bf Q+q})=V_0+\sum_{i=y,z}V_i\cos(q_i \delta_i),
\end{equation} 
the pinning potential is obtained as
\begin{eqnarray}
V_{imp}(\Phi)=-\frac{8V_0V_yN_0^2}{\pi}\sum_j\cos(2({\bf QR}_j+\Phi({\bf R}_j)))
\epsilon_0\int_{0}^{a+1}\tanh\frac{\epsilon_0 x}{2T}\frac{AB}{\pi^2}dx,
\end{eqnarray}
where
\begin{eqnarray}
A=\int\limits_{\gamma}^{\delta}E\left(\sqrt{1-\left(\frac{x
-\cos(y)}{a}\right)^2}\right)dy,\\
B=\int\limits_{\gamma}^{\delta}\textmd{sgn}(x-\cos(y))\left[K\left(
\frac{x-\cos(y)}{a}
\right)-E\left(\frac{x-\cos(y)}{a}\right)\right]dy+\nonumber \\
+\Theta(\pi-\delta)\int\limits_{\delta}^\pi \frac{|x-\cos(y)|}{a} 
\left[K\left(\frac{a}{x-\cos(y)}\right)-E\left(\frac{a}{x-\cos(y)}\right)
\right]dy+
\nonumber \\
+\Theta(\gamma)\int\limits_{0}^{\gamma} \frac{|x-\cos(y)|}{a}
\left[K\left(\frac{a}{x-\cos(y)}\right)-E\left(\frac{a}{x-\cos(y)}\right)
\right]dy
\end{eqnarray}
\begin{eqnarray}
\gamma=\arccos\left(1-2\textmd{max}\left(0,\frac{1-a-x}{2}\right)\right),\\
\delta=\arccos\left(1-2\textmd{min}\left(1,\frac{a+1-x}{2}\right)\right),
\end{eqnarray}
$a=\Delta/\epsilon_0$.
Then following FLR\cite{fl,lr}, the threshold electric field  in the 
strong pinning limit reads as
\begin{equation}
\frac{E_T^S(T,\epsilon_0)}{E_T^S(0,0)}=\frac{\rho_s(0,0)}{\rho_s(T,\epsilon_0)}
\frac{\epsilon_0}{0.5925}\int_{0}^{a+1} \tanh\frac{\epsilon_0x}{2T}
\frac{AB}{\pi^2}dx.
\end{equation}
For any finite $\epsilon_0$ the
density of states has no zero at the Fermi energy (Eq. (\ref{eq:res3d})), and
the threshold electric field increases as
\begin{gather}
\frac{E_T^S(T,h)}{E_T^S(0,0)}=\left\{
\begin{array}{lc}
1+2\ln(2) \dfrac{T}{\Delta_{00}} & \epsilon_0\ll T\\
1+\dfrac{2\epsilon_0}{\pi\Delta_{00}} & T\ll \epsilon_0
\end{array}\right.
\end{gather}
in the $\epsilon_0,T\ll \Delta_{00}$ range.
The threshold electric field along the second order phase 
boundary is obtained as
\begin{equation}
\frac{E_T^S(T,\epsilon_0)}{E_T^S(0,0)}=-\frac{\tanh\dfrac{\epsilon_0}{2T}}
{\left\langle\Psi^{\prime\prime}\left(\dfrac12+\dfrac{i\epsilon_0\cos{y}}{2\pi
T}\right)\right\rangle}\frac{T^2}{\Delta_{00}\epsilon_0}\frac{\pi^5}{4\times 
0.5925}.
\end{equation}
The threshold electric field is shown in Fig. \ref{fig:et3d}.
Close to the critical nesting where the transition temperature tends to zero, 
threshold electric field diverges at the actual transition temperature as 
\begin{eqnarray}
\frac{E_T^S(T,\epsilon_0)}{E_T^S(0,0)}=4.18\frac{T^2}{\Delta_{00}\epsilon_0}
\exp\left(1.03\frac{\epsilon_0}{T}\right).
\end{eqnarray}

\begin{figure}[h!]
\psfrag{x}[t][b][1][0]{$T/T_{c}$}
\psfrag{y}[b][t][1][0]{$E_T^W(mV/cm)$}
{\includegraphics[width=13cm,height=8cm]{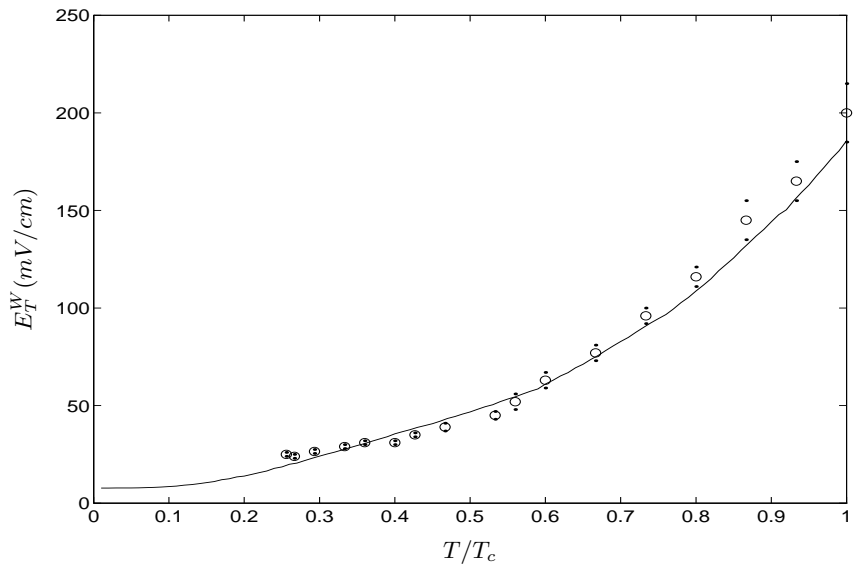}}
\caption{The threshold electric field in the weak pinning limit is plotted as 
a function of 
reduced temperature for $\epsilon_0=0.8\Delta_{00}=0$ 
 together with the experimental data (open circles). The dots represent
the error bar. \label{fig:et8boj}} 
\end{figure}

The strong pinning limit applies when
impurities are introduced by X-ray irradiation or some violent means. In
this case one single impurity is enough to pin the
DW locally. For high quality crystals, however, the weak pinning limit 
is more appropriate\cite{tmtsf}. 
Then for the 3 dimensional weak pinning limit we obtain\cite{viro1,viro2}
\begin{equation}
\frac{E_T^W(T,\epsilon_0)}{E_T^W(0,0)}=\left(\frac{E_T^S(T,\epsilon_0)}{E_T^S(0,0)}\right)^4.
\end{equation}
$E_T^W(T,\epsilon_0)$ is plotted for $\epsilon_0=0.8\Delta_{00}$ as a function of 
temperature together with the
experimental data on $\alpha$-(ET)$_2$KHg(SCN)$_4$\cite{ltp} 
in Fig. \ref{fig:et8boj}. 
The fit was made with the least square method. 
As we have foreseen\cite{rapid}, imperfect nesting has improved the agreement
remarkably well. The present theoretical results apply also for USDW.

The agreement between the experimental and theoretical threshold electric field 
data, together with the magnetic field-temperature phase 
diagram\cite{tesla,jetp}, the specific
heat and the lack of any spatially periodic density modulation suggests, 
that the
LTP of the $\alpha$-(BEDT-TTF)$_2$KHg(SCN)$_4$ salt is probably UCDW.

\section{Conclusion}

We have studied theoretically the effect of imperfect nesting in unconventional density
waves. 
We explored the phase diagram which is identical to the one
in conventional density wave. 
In the density of states, the peaks at $\pm\Delta$ of the
perfectly nested system split into cusps at $\pm\Delta\pm\epsilon_0$. 
Usually $\epsilon_0$ is thought to vary with
pressure providing the opportunity to check these result in a wide range of
parameters. For example, the density of states obtained from ARPES measurements at
different applied pressures can be compared to our predictions.
Only the low frequency part of the conductivity is
chopped transferring spectral weight to the Dirac delta peak at $\omega=0$
(similarly to the effect of magnetic field\cite{tesla}),
which is expected to be broadened due to impurity effects. 
The threshold field shows
similar features to the one in conventional DW, increases smoothly with
$\epsilon_0$ allowing us to describe the the measured $E_T$ of
$\alpha-$(ET)$_2$ salts\cite{ltp} with
$\epsilon_0=0.8\Delta_{00}$. 
This supports
our proposal that the LTP of $\alpha-$(ET)$_2$ salts is an unconventional charge
density wave.

\begin{acknowledgments}
We thank Bojana Korin-Hamzi\'c for sending us the experimental data before
publication which provided us with initial stimulus to undertake this work. 
We thank also
Mark Kartsovnik, Takahiko Sasaki and Peter Thalmeier for useful discussions. 
This work
was supported by the Hungarian National Research Fund under grant numbers
OTKA T032162 and T037451, and by the Ministry of Education under grant number
FKFP 0029/1999.
\end{acknowledgments}

\bibliography{eth}
\end{document}